\documentclass[12pt]{article}

\usepackage[dvips]{graphicx,color}

\begin{document}

\begin{center}

{\Large {\bf Potential for the slow-roll inflation,  \\

\vspace{0,4cm}

mass scale hierarchy and Dark Energy  \\

\vspace{0,5cm}
  
from the Type IIA supergravity}}

\vspace{1,5cm}

{Boris L. Altshuler}\footnote[1]{E-mail adresses: altshuler@mtu-net.ru \& altshul@lpi.ru}

\vspace{0,5cm}

{\it Theoretical Physics Department, P.N. Lebedev Physical
Institute, \\  53 Leninsky Prospect, Moscow, 119991, Russia}

\vspace{1,5cm}

{\bf Abstract}

\end{center}

The magnetic fluxbrane solution with a strongly warped throat is studied in 
the Type IIA supergravity theory with co-dimension one local source which 
is the $Z_{2}$-symmetric UV boundary of the throat. Overall volume of extra 
space may be stabilized since introduction of the local source breaks the no-scale 
structure of the theory and evades the no-go theorem. Radion field is defined 
as the position of UV boundary "moved" from its stable value fixed by the 
anisotropic Israel junction conditions. Analytical expression for the radion 
effective potential is received. Potential decreases 
exponentially (exponent is equal to 0,21 in Planck units) in the slow-roll 
region and apparently meets other demands of the early inflation. Thus radion 
surves an inflaton. Reissner-Nordstrom type deformation of 
the elementary fluxbrane solution 
permits to construct the IR end of the throat and results in tiny positive 
non-zero value of the radion potential in its extremal point seen today as 
Dark Energy $\rho_{D.E.}$. Expressions for the mass scale 
hierarchy $m/M_{\rm Pl}$ and for the "acceleration hierarchy" received 
in the paper give the physically interesting 
relation between two hierarchies: $\rho_{D.E.} \sim m^{8}/M_{\rm Pl}^{4}$.

\vspace{0,5cm}

\newpage
\section{Introduction}

\qquad This paper continues the research of \cite{Alt06} where 
analytical expression for the radion effective potential capable to 
meet demands of the slow-roll inflation was received in a class of models 
with fluxbrane throat-like solutions. These studies develop the direction 
of thought of papers \cite{Strominger}-\cite{Wolfe2} where the throat-like 
solutions in the Type IIB supergravity with warped Klebanov-Strassler 
conifold \cite{Klebanov}-\cite{Herzog} were considered. In the present paper 
we consider the non-deformed and deformed throat-like solution in the 
Type IIA supergravity where six extra dimensions are given by the warped 
flat space which base is a sphere (\cite{Horowitz}-\cite{Aharony}). 

Following the conventional approach, and in parallel with the Randall-Sundrum 
theory \cite{Randall}, we suppose that massive matter of Standard 
Model (SM) is localized 
in the IR region near the tip of the throat whereas the volume of extra 
space is terminated by the co-dimension one local source - heavy "ultraviolet" 
boundary where $Z_{2}$-identification and corresponding junction 
conditions are 
imposed. And we suppose that dynamics of the boundary 
surface is described by the simplest positive tension Nambu-Goto action, 
hence energy-momentum tensor of the boundary is isotropic. In Sec. 3 it 
is shown that anisotropic Israel junction conditions stabilize position of 
the isotropic boundary at the top of the throat. 

This seemingly comes in conflict with the no-scale structure of 
the supergravity theory (\cite{Giddings} and references therein) and with 
the no-go theorem \cite{nogo}, \cite{Giddings}. However there is no conflict 
at all. To fix the modulus of the overall volume of extra space is possible 
because the no-scale structure is spoiled here by the external local source 
(UV boundary), and because this co-dimension one 
local source also evades the no-go theorem. Indeed 
it is easy to show that combination $\widetilde T$ (defined 
by expression (34) of the paper \cite{nogo}) of 
the components of the energy-momentum tensor does not meet demands of 
the no-go theorem in case the positive tension
co-dimension one local source is introduced in the action; this is not 
true however for the positive tension local sources of lower dimensions.

The Reissner-Nordstrom type deformation of the elementary extremal solution 
will be used as a tool to construct the IR end of the throat. This type of 
solutions with a "bolt" (in terminology 
of \cite{Hawking}) was considered earlier in 6D models 
\cite{Louko}-\cite{Altsh1} where also the constant curvature of the 
4-dimensional space-time was introduced to make Israel junction conditions 
consistent. As it was shown in \cite{Altsh1} the value of this curvature must 
be extremely small and may correspond to the observed acceleration of 
the Universe (see Review \cite{quint}). In the present paper this approach is 
generalized in D10 Type IIA supergravity theory. Contrary to \cite{Altsh1} 
where formula for Dark Energy $\rho_{D.E.}=G_{N}m^{6}$ ($G_{N}$ is Newton's 
constant, $m$ is characteristic mass of matter) was received, model considered 
in the present paper gives more realistic 
result $\rho_{D.E.} \sim G_{N}^{2}m^{8}$ (subsection {\it 5-b}).

Radion field is defined here as the position of the UV boundary which 
slowly depends on the space-time coordinates in 4 dimensions 
(cf. \cite{radion3}-\cite{Brax}). Radion effective potential 
is calculated with the standard procedure of 
integrating out extra coordinates in the higher-dimensional action. 
Of course it is possible to rescale the isotropic radial coordinate and to 
make the position of the boundary fixed; then we come to the definition of 
radion as a factor-field in higher-dimensional 
metric \cite{Csaki} - \cite{Mazumdar}. Two approaches are basically 
equivalent when radion effective potential is calculated.

Form of the radion potential depends only on the choice of the theory; in 
the Type IIA supergravity considered in the paper potential exponentially 
decreases upward the throat with an exponent equal to 0,21, i.e. it is 
sufficiently flat to provide the slow-roll inflation \cite{Dvali}, \cite{Mukhanov}. 
Thus radion scalar field introduced in the paper may serve as an inflaton. 
Radion potential proves to be non-negative, its relatively flat 
region ends with steep slope falling down to the minimum of potential at the 
top of the throat. In the present paper these results of \cite{Alt06} are 
repeated in a more transparent way. Also the validity of the hypothesis 
of \cite{Alt06} is proved: it is shown in subsection {\it 5-c} 
that Reissner-Nordstrom 
type deformation of the elementary fluxbrane solution results in tiny 
positive deviation from zero of the minimal value of the radion effective 
potential. This deviation is seen today as Dark Energy.

The idea to use dynamical scalar associated with extra dimensions, 
interbrane distance in particular, as a candidate for inflaton is not a 
novel one (see e.g. \cite[b]{Cline}, \cite{Mazumdar}). The very 
possibility to get in this way the exact analytical expression for the 
scalar field potential possessing qualitatively the basic features 
demanded by the astrophysical observations \cite{WMAP} looks attractive. 

It must be noted that physically meaningful radion effective potential 
may be received here only in case the bulk magnetic monopole fluxbrane 
solution is considered as a background, not the dual electric one. The nonequivalence of 
two solutions is immediately seen when the higher-dimensional consistency 
condition of \cite{Leblond} is applied, see Appendix in \cite{Alt06}.

The formulae for the value of the electro-weak hierarchy presented in 
Sec. 5 essentially develop ideas of works \cite{Altsh} where it was observed 
that in the throat-like fluxbrane models hierarchy proves to be 
strongly dependent on the value of the $n$-form-dilaton coupling constant 
and on dimensionalities of the extra subspaces.

The paper is organized as follows. Basic action, bulk and junction 
equations are presented in Sec. 2. In Sec. 3 the stabilization of the 
volume modulus of the non-deformed throat-like 
solution in the Type IIA supergravity is demonstrated, the 
analytical expression for the radion effective potential is received and its 
compatibility with the demands of inflation is shown. In Sec. 4 the 
generalization of the elementary solution induced 
by introduction of non-zero "Maxwell" field and non-zero curvature 
of the 4-dimensional Universe is considered. Sec. 5 presents formulae 
for the mass-scale hierarchy and for the rate of 
acceleration of the Universe; the physically meaningful dependence of 
two hierarchies is deduced. 
In Sec. 6 results and problems are summarized and possible trends 
of future research are outlined.

\section{Action, ansatz, dynamical equations}

Let us consider the following action in $D$ dimensions:

\begin{eqnarray}
\label{1}
&&S^{(D)}=M^{D-2}\Bigg \{ \int\left[R^{(D)}-\frac{1}{2}(\nabla\varphi)^2-\frac{1}{2\cdot n!}e^{\alpha\varphi}F_{(n)}^2-\frac{1}{2\cdot 2!}e^{\eta\varphi}F_{(2)}^2\right.-  \nonumber   
\\
&&-\left.\sigma \, e^{\gamma\varphi} \delta^{(1)}\frac{\sqrt{-h^{(D-1)}}}{\sqrt{-g^{(D)}}}\right]\sqrt{-g^{(D)}}\,d^{D}x+\rm{GH} \Bigg \},
\end{eqnarray}
which bulk part is an Einstein-frame truncated low-energy description of the string-based supergravity with dilaton and antisymmetric tensor;
$M$, $g_{AB}$, $R^{(D)}$ are "Planck mass", metric and curvature in $D$ dimensions; $\rm{GH}$ - Gibbons-Hawking 
term; $F_{(n)}$ is $n$-form field strength; $F_{(2)}$ is 2-form "Maxwell" field; $\varphi$ - dilaton field coupled to $n$-form, 2-form and to local source in (\ref{1}) with a coupling constants $\alpha$, $\eta$, $\gamma$ 
correspondingly. The co-dimension one local source will serve the UV-boundary of the throat, its action is taken in the simplest Nambu-Goto form; mass parameter $\sigma$ characterises its tension equal to $M^{D-2}\sigma$; 
${h^{(D-1)}=\det{h_{ab}}}$; $h_{ab}$ is an induced metric on the boundary, $a, \,b=\{0,1\dots(D-2)\}$; $\delta^{(1)}$ is Dirac delta function fixing position of the boundary.

In this paper we shall consider theory (\ref{1}) for the following particular values of dimensionalities and coupling constants in (\ref{1}):

\begin{equation}
\label{2}
D=10, \qquad n=4, \qquad \alpha=\frac{1}{2}, \qquad \eta=\frac{3}{2}, \qquad \gamma=-\frac{1}{12}.
\end{equation}
With this choice the bulk part of the action (\ref{1}) is a truncated Bose-action of 
the Type IIA supergravity. It is worthwile to note that D10 
theory (\ref{1}) with specific values of constants given in (\ref{2}) is
just a compactification of the action of D11 $M$-theory where in addition to the conventional bulk terms the D10 local source is included:

\begin{equation}
\label{3}
S_{(M)}=\tilde{M}^{9}\Bigg \{ \int\left[R^{(11)}-\frac{1}{2\cdot 4!}F_{(4)}^{2}-\tilde{\sigma}\delta^{(1)}\frac{\sqrt{-h^{(10)}}}{\sqrt{-g^{(11)}}}\right]\sqrt{-g^{(11)}}\,d^{11}x+\rm{GH} \Bigg \},
\end{equation}
$\tilde{M}$, $\tilde{\sigma}$ are Planck mass and mass parameter of the local source in 11 dimensions. After reduction of action (\ref{3}) to 10 dimensions the volume of compact 11-th dimension becomes a dilaton field whereas 2-form in (\ref{1}) is a corresponding Kaluza-Klein field. However we shall not refer any more to $M$-theory and consider the supergravity action (\ref{1}), where dimensionalities and constants are given in (\ref{2}), as a primary one throughout the paper.

The following ansatz for the bulk 
solution of the dynamical equations given by the action (\ref{1}), (\ref{2}) will be used:

\begin{eqnarray}
\label{4}
&&ds_{(10)}^{2}=b^{2}{\tilde g}_{\mu\nu}dx^{\mu}dx^{\nu}+ c^{2}dz^{2}+N^{2}dr^{2}+a^{2}d\Omega_{4}^{2}, \qquad \varphi=\varphi(r),
\\
&& F_{(4)}=Q_{(4)}dy^{1}\wedge dy^{2}\wedge dy^{3}\wedge dy^{4}, \, \,   F_{(2)zr}=dA_{z}(r)/dr=\frac{Q_{(2)}cN}{b^{4}a^{4}}\, e^{-3\varphi /2}, \nonumber
\end{eqnarray}
where metric scale factors $b$, $c$, $a$, "lapse function" $N$ and dilaton $\varphi$
depend only on the isotropic coordinate $r$, ${\tilde g}_{\mu\nu}$ is 
metric of the 4-dimensional Universe $M_{(3+1)}$, $z$ is coordinate of 
torus $S^{1}$ of period $T_{z}$, $d\Omega_{4}^{2}$ is metric of 4-sphere of unit 
radius; $x^{A}=\{x^{\mu},z,r,y^{i}\}$, $A=0,1\ldots 9$, $\mu=0,1,2,3$, $i=1,2,3,4$.
$Q_{(4)}$ is charge of the magnetic monopole. $A_{z}$ is non-zero component of 
the vector-potential of 2-form field $F_{(2)}$, $Q_{(2)}$ is its "electric" charge, 
last equality for $F_{(2)}$ in (\ref{4}) is received from the "Maxwell" 
equation for 2-form written down for the metric ansatz (\ref{4}). 

Introduction of small $F_{(2)}\ne 0$ gives 
the Euclidian version of the Reissner-Nordstrom type deformation 
of the extremal fluxbrane solution. It 
will be shown in Sec. 4 that this deformation provides the physical tool to terminate 
the throat at its IR end and also enforces dynamically to introduce the 
extremely small positive curvature ${\widetilde R}^{(4)}=12{\tilde h}^{2}$
of the manifold $M_{(3+1)}$; auxiliary "Hubble constant" $\tilde h$ is 
connected with the observed acceleration rate of the 
Universe $h=10^{-60} M_{\rm Pl}$ by scale 
transformation (\ref{30}) below (see in Sec. 5 in more detail).

With ansatz (\ref{4}) and ${\tilde h}\ne 0$ action (\ref{1}) with 
parameters (\ref{2}) in it gives following gravity equations for scale 
factors $b(r)$, $c(r)$, $a(r)$ (we don't need to put down the 
gravity constraint) and equation for dilaton field (prime means 
derivative over $r$):

\begin{equation}
\label{5}
\frac{3{\tilde h}^{2}}{b^{2}}+\frac{1}{N^2}\Bigg[-\frac{b''}{b}+\frac{b'^{2}}{b^{2}}+\frac{b'}{b}\Bigg(\frac{N'}{N}-4\frac{b'}{b}-\frac{c'}{c}-4\frac{a'}{a}\Bigg)\Bigg]=-\frac{3}{8}J_{(4)}-\frac{1}{8}J_{(2)}+\frac{1}{16}J_{(\sigma)},
\end{equation}
\\
\begin{equation}
\label{6}
\frac{1}{N^2}\Bigg[-\frac{c''}{c}+\frac{c'^{2}}{c^{2}}+\frac{c'}{c}\Bigg(\frac{N'}{N}-4\frac{b'}{b}-\frac{c'}{c}-4\frac{a'}{a}\Bigg)\Bigg]=-\frac{3}{8}J_{(4)}+\frac{7}{8}J_{(2)}+\frac{1}{16}J_{(\sigma)},
\end{equation}
\\
\begin{equation}
\label{7}
\frac{3}{a^{2}}+\frac{1}{N^2}\Bigg[-\frac{a''}{a}+\frac{a'^{2}}{a^{2}}+\frac{a'}{a}\Bigg(\frac{N'}{N}-4\frac{b'}{b}-\frac{c'}{c}-4\frac{a'}{a}\Bigg)\Bigg]=-\frac{5}{8}J_{(4)}-\frac{1}{8}J_{(2)}+\frac{1}{16}J_{(\sigma)},
\end{equation}

\begin{equation}
\label{8}
\frac{1}{N^2}\Bigg[\varphi''-\varphi'\Bigg(\frac{N'}{N}-4\frac{b'}{b}-\frac{c'}{c}-4\frac{a'}{a}\Bigg)\Bigg]=\frac{1}{2}J_{(4)}+\frac{3}{2}J_{(2)}-\frac{1}{12}J_{(\sigma)},
\end{equation}

where

\begin{eqnarray}
\label{9}
&&J_{(4)}\equiv \frac{e^{\varphi /2}F_{(4)}^{2}}{2\cdot 4!}=\frac{e^{\varphi /2}Q_{(4)}^{2}}{2a^{8}}, \qquad   J_{(2)}\equiv \frac{e^{3\varphi /2}F_{(2)}^{2}}{2\cdot 2!}=\frac{e^{-3\varphi /2}Q_{(2)}^{2}}{2b^{8}a^{8}}, \nonumber \\
\\
&&J_{(\sigma)}\equiv e^{-\varphi /12} \, \sigma \frac{\delta(r-r_{0})}{N},  \qquad {\tilde h}^{2}=\frac{{\widetilde R}^{(4)}}{12}. \nonumber
\end{eqnarray}

We suppose that local source is placed at some $r=r_{0}$ and that it 
terminates the throat "from above", i.e. it forms the UV end of the throat. Thus we truncate space-time (\ref{4}) at $r=r_{0}$, paste two copies of 
the inner region along the cutting surface and consider codimension one local source 
in the RHS of equations (\ref{5})-(\ref{8}) as a heavy boundary 
where $Z_{2}$-symmetry is imposed. The space-time of the surface is a product

\begin{equation}
\label{10}
M_{(3+1)} \times S^{1} \times S^{4},
\end{equation}
physically this boundary may be or 8-brane which spans $M_{(3+1)}$ and 
wraps around compact extra spaces $S^{1}$, $S^{4}$, or it may be
viewed as a shell of branes of lower dimension uniformly distributed over 
compact subspaces. In any case we suppose, and this is the main 
hypothesis of the paper, that dynamics of the boundary surface is described 
by the simplest Nambu-Goto action, hence its energy-momentum tensor 
is isotropic as it follows from the action (\ref{1}). There are four junction 
conditions at the boundary: three Israel conditions for three subspaces 
in (\ref{10}) and jump condition for dilaton field $\varphi$. These 
conditions are immediately received by integrating 
equations (\ref{5})-(\ref{8}) over $r$ around $r=r_{0}$ (factor 2 in the 
LHS in (\ref{11})-(\ref{14}) reflects the $Z_{2}$-symmetry):

\begin{equation}
\label{11}
\frac{2}{N^{2}} \, \frac{b'}{b}=\frac{\sigma e^{-\varphi/12}}{16N},
\end{equation}

\begin{equation}
\label{12}
\frac{2}{N^{2}} \, \frac{c'}{c}=\frac{\sigma e^{-\varphi/12}}{16N},
\end{equation}

\begin{equation}
\label{13}
\frac{2}{N^{2}} \, \frac{a'}{a}=\frac{\sigma e^{-\varphi/12}}{16N},
\end{equation}

\begin{equation}
\label{14}
-\frac{2}{N^{2}} \, \varphi'=-\frac{1}{12} \frac{\sigma e^{-\varphi/12}}{N}.
\end{equation}

These relations must be valid at the position of the boundary $r=r_{0}$. 
Equations (\ref{11})-(\ref{14}) are actually quite 
informative. We'll see that they fix position $r_{0}$ of the UV boundary, i.e. 
determine the overall volume of the extra space, they "fine-tune" 
magnetic monopole charge $Q_{(4)}$ and mass parameter $\sigma$ in (\ref{1}), 
and in the model with deformed extremal
solution considered in Sec. 4 the consistency of equations (\ref{11}) and (\ref{12}) demands 
introduction of non-zero curvature  of 
the Universe which value, as it will be shown, may correspond to the 
observed acceleration of the Universe.

\section{Non-deformed fluxbrane solution, stabilization of the volume modulus and potential for the slow-roll inflation}

\vspace{0,5 cm}

{\large\it 3-a. Bulk solution and stabilization of the volume modulus.}

\vspace{0,5cm}

In this Section the way of thought of paper \cite{Alt06} is repeated for the 
particular case of the Type IIA supergravity given by action (\ref{1}) with constants (\ref{2}) in it
where we discard "Maxwell" field and consider $M_{(3+1)}$ the Minkowsky 
space-time (i.e. put $F_{(2)}=0$, ${\tilde h}=0$ in (\ref{5})-(\ref{8})). 
Then ansatz (\ref{4}) for the elementary magnetic fluxbrane solution 
of the equations (\ref{5})-(\ref{8}) looks as \cite{Gibbons}- \cite{Aharony}:

\begin{eqnarray}
\label{15}
&&ds_{(10)}^{2}=H^{-3/8}({\tilde g}_{\mu\nu}dx^{\mu}dx^{\nu}+ dz^{2})+H^{5/8}(dr^{2}+r^{2}d\Omega_{4}^{2}), \quad e^{\varphi}=e^{\varphi_{\infty}}H^{-1/4}, \nonumber \\
\\
&& F_{(4)}=Q_{(4)}dy^{1}\wedge dy^{2}\wedge dy^{3}\wedge dy^{4}, \quad  H=1+\left (\frac{L}{r}\right )^{3}, \quad L^{3}=\frac{1}{3}Q_{(4)}e^{\varphi_{\infty}/4}, \nonumber
\end{eqnarray}
$\varphi_{\infty}$ is the value of dilaton field at $r=\infty$.

Metric (\ref{15}) describes a warped "throat" of proper length 
$\int_{0}^{L}H^{5/16}dr\cong 16L$ with an integrable singularity 
at $r=0$ where curvature $R^{(10)}\to \infty$. 
The low-energy action (\ref{1}) makes sense only if curvature components are 
small as compared to $M^{2}$. For the space-time (\ref{15})
this condition formulated e.g. for the scalar curvature in 10 dimensions reads:

\begin{equation}
\label{16}
R^{(10)}= \frac{45}{32}\, \frac{1}{L^{2}}\left (\frac{L}{r}\right )^{1/8}< M^{2},
\end{equation}
this inequality is written inside the throat where $r \ll L$.
                                  
From (\ref{16}) the minimal permitted value $r_{\it{min}}$ of the isotropic 
coordinate is immediately determined:

\begin{equation}
\label{17}
r>r_{\it{min}}= k^{8}\, (ML)^{-16} \, L,
\end{equation}
where coefficient $k$ is equal $45/32$ when $R^{(10)}$ is used in the 
estimate inequality (\ref{16}). In what follows we shall consider $k$ 
some number of order one.

Big value of the exponent in the RHS of (\ref{17}) reflects its non-analytical 
dependence on the $4$-form-dilaton coupling constant $\alpha$. In general case this 
exponent is equal to $\Delta/ \alpha^{2}$ \cite{Alt06} \footnote[1]{There is 
unfortunate mistake in \cite{Alt06} where factor $2$ is omitted in the 
denominator of exponent in the RHS of expression (\ref{14}) of \cite{Alt06}; 
this however does not change essentially conclusions of \cite{Alt06}. 
The present paper gives the corrected formulae when the Type IIA supergravity 
is considered.} ($\Delta$ is well known 
parameter of the elementary fluxbrane solutions determined by $\alpha$ and 
dimensionalities; $\Delta=4$ in the Type IIA, Type IIB supergravities).
For $\alpha=0$ in (\ref{1}), like it takes place in the Type IIB supergravity 
where there is $AdS_{5}\times S^{5}$ asymptotic inside the throat, 
$r_{\it{min}}=0$, there is no singularity of curvature at any $r$.

Comparison of metrics (\ref{15}) and (\ref{4}) gives $b=c=H^{-3/16}$, 
$a=H^{5/16}\, r$, hence jump conditions (\ref{11}), (\ref{12}) coincide; also 
dilaton jump condition (\ref{14}) identically follows from (\ref{11}) for dilaton 
field given in (\ref{15}). Compatibility of jump conditions(\ref{11}) 
and (\ref{13}) demands 
$b'/b=a'/a$, i.e. $(Hr^{2})'=0$ at $r=r_{0}$ which gives:

\begin{equation}
\label{18}
r_{0}=\frac{L}{2^{1/3}},
\end{equation}
whereas (\ref{11}) with account of (\ref{18}) connects $L$, $\sigma$ and $\varphi_{\infty}$. We shall put down this 
relation multiplying it by the higher-dimensional Planck mass $M$:

\begin{equation}
\label{19}
ML=12\left (\frac{2}{3}\right )^{1/3}\frac{M}{\sigma} \, e^{\varphi_{\infty}/12}\equiv 10,5 \, g,
\end{equation}
where dimensionless constant

\begin{equation}
\label{20}
g=\frac{M}{\sigma} \, e^{\varphi_{\infty}/12}
\end{equation}
is an important parameter which determines the physical predictions of the 
model. $g$ is an invariant of scale transformation 
$g_{AB} \to e^{2\lambda}g_{AB}$, $\varphi \to \varphi + 12\lambda$, 
$M \to e^{-\lambda}M$ ($\lambda=const$) which is an unvariance of the action 
(\ref{1}) when $F_{(2)}=0$ and constants (\ref{2}) are used in (\ref{1}).
All the approach makes sense only if $ML\gg 1$, hence it is necessary 
that $g \ge 1$ as it follows from (\ref{19}).

From (\ref{19}), (\ref{15}) it also follows the "fine-tuning" condition:

\begin{equation}
\label{21}
Q_{(4)}^{1/3}=12\cdot 2^{1/3}\cdot \sigma^{-1}.
\end{equation}
This is a direct analogy of the fine-tuning of the bulk cosmological constant and brane's tension demanded in the Randall-Sundrum model \cite{Randall}. However here the bulk magnetic $4$-form charge $Q_{(4)}$ is not an input parameter in the action but a free constant of the bulk solution of the dynamical equations. Hence relation (\ref{21}) is by no means a fine-tuning but a constraint determining magnetic charge through the constant $\sigma$ of the action (\ref{1}).

We'll see below that position (\ref{18}) of the UV boundary of the throat determining the overall volume of extra space is a point of zero minimum of the corresponding effective potential. As it was already noted in the Introduction dynamical stabilization of the modulus of the volume of extra space is possible because in the model under consideration the local source in (\ref{1}) breaks the no-scale structure of the theory and because this codimension one local source also violates the conditions of the no-go theorem.

Thus dynamics of the model terminates the extra space of the space-time (\ref{15}) at the top of the throat, at its UV end (\ref{18}). In Sec. 4 the IR end of the throat where supposedly Standard Model resides will be fixed near the tip of the throat by the tool of small deformation of the extremal solution (\ref{15}). This deformation will not influence essentially the form of the radion effective potential calculated in the next Subsection for the non-deformed background (\ref{15}), but will result in a tiny positive deviation (seen today as Dark energy) from the minimal zero value of potential received for the non-deformed background (see Sec. {\it {5-c}}).

\vspace{0,5 cm}

{\large\it 3-b. Radion as inflaton. Potential for the slow-roll inflation.}

\vspace{0,5cm}

Effective action $S^{(3+1)}$ in four dimensions is conventionally calculated by integrating out extra coordinates in a higher dimensional action. With account of the radion field this will give the effective scalar-tensor Brans-Dicke type action (see e.g. \cite[b]{Kanno}, \cite{Csaki}). To calculate the effective action $S^{(3+1)}$ we shall use in (\ref{1}) the bulk solution (\ref{15}) but move the UV boundary (and hence change the upper limit of the integration over isotropic coordinate $r$ in (\ref{1})) from $r=r_{0}$ (\ref{18}) fixed by junction conditions (\ref{11})-(\ref{14}) to the arbitrary position $\rho (x)$, slowly depending on coordinates $x^{\mu}$:

\begin{equation}
\label{22}
r_{0} \to \rho (x),
\end{equation}
$\rho (x)$ is called radion field \cite[b]{Cline}, \cite{radion3}-\cite{Brax}. This definition of the radion field is equivalent to the more conventional one where radion is considered as depending on $x^{\mu}$ factor of the lapse-function $N$ in metric (\ref{4}) \cite{Csaki}-\cite{Mazumdar}. The gradient terms of $\rho (x)$ contribute to the induced metric of the UV boundary:

\begin{equation}
\label{23}
h_{ab}=g_{ab}+\rho,_{a}\rho,_{b}g_{rr},
\end{equation}
where $x^{a}=\{x^{\mu}, z, y^{i}\}$ and $g_{ab}$, $g_{rr}$ are corresponding components of the bulk metric (\ref{15}). Then, considering that $\rho(x)$ does not depend on $z$, $y^{i}$ and depends on $x^{\mu}$ slowly as compared to the scales of the bulk solution, the Lagrangian $L_{\it{l.s.}}$ of the local source in (\ref{1}) (with constants (\ref{2}) in it and with account of (\ref{15})) takes the form:

\begin{eqnarray}
\label{24}
&&{\rm L}_{\it{l.s.}}=-M^{8}\sigma e^{-\varphi /12}\delta(r-\rho)\,\frac{\sqrt{-h^{(9)}}}{\sqrt{-g^{(10)}}} \approx  \nonumber
\\
&&\approx -\frac{M^{8}\sigma e^{-\varphi_{\infty}/12}\,\delta(r-\rho)}{H^{14/48}}\left[1+\frac{1}{2}H {\tilde g}^{\mu\nu}\rho,_{\mu}\rho,_{\nu}\right].
\end{eqnarray}

In calculating radion effective potential we shall substitute $r_{\it IR} \to r=0$ in the lower limit of the integration over $r$ in (\ref{1}). This will not change $S^{(3+1)}$ essentially since it is supposed that $r_{\it IR}\ll L$ and since all integrals in (\ref{1}) are convergent at $r=0$. We also postulate that $Z_{2}$-symmetry at the "moved" UV boundary surface is preserved, i.e. bulk integration must be fulfilled over two pasted copies of the solution. Direct calculation shows that for the magnetic monopole fluxbrane bulk solution (\ref{15}) this procedure gives zero value of the radion potential at $\rho=r_{0}$ (\ref{18}) where jump conditions (\ref{11})-(\ref{14}) are valid; this corresponds to the consistency conditions \cite{Leblond}. And this is not the case for the dual bulk electric fluxbrane solution - see {\it {Note 2}} below.

Thus symbolically the Brans-Dicke type effective action $S^{(3+1)}$ depending on general meric ${\tilde g}_{\mu\nu}(x)$ of the manifold $M_{(3+1)}$ and on the radion field $\rho (x)$ is received when extra coordinates are integrated out in the action (\ref{1}):

\begin{eqnarray}
\label{25}
&&S^{(4)}=2\int_{0}^{\rho}{\rm L}_{\it bulk}+\int {\rm L}_{\it{l.s.}}=
\\
&&=\int\left[\Phi(\rho){\widetilde R}^{(4)}-\frac{1}{2}\omega (\rho){\tilde g}^{\mu\nu}\rho,_{\mu}\rho,_{\nu}-{\widetilde V}(\rho)\right]\sqrt{-{\tilde g}^{(4)}}d^{(4)}x, \nonumber
\end{eqnarray}
where $L_{\it bulk}$ sums up all bulk terms in (\ref{1}) including the Gibbons-Hawking term, $L_{\it{l.s.}}$ is given in (\ref{24}); ${\widetilde R}^{(4)}$ is scalar curvature of the $(3+1)$ dimensional space-time described by metric ${\tilde g}_{\mu\nu}(x)$ slowly depending on $x^{\mu}$. Brans-Dicke field $\Phi (\rho)$, kinetic term function $\omega (\rho)$ and auxiliary radion potential ${\widetilde V}(\rho)$ in (\ref{25}) are calculated when bulk metric (\ref{15}) is used in (\ref{1}) where it is taken ${\tilde g}_{\mu\nu}=\eta_{\mu\nu}$ - Minkowski metric in four dimensions; $Q_{(4)}$, $\sigma$, $\varphi_{\infty}$ may be expressed through the characteristic length of the throat $L$ with use of dependences given in (\ref{15}), (\ref{19}); $T_{z}$ is period of torus $S^{1}$ in (\ref{15}), $\Omega_{4}$ is volume of four-sphere of unit radius. Simple calculations finally give:

For the Brans-Dicke field:

\begin{equation}
\label{26}
\Phi(\rho)=2M^{8}\Omega_{4} T_{z} \int_{0}^{\rho} Hr^{4}\,dr \, = \, 2M^{2}(ML)^{5}(MT_{z})\Omega_{4} \left[\frac{1}{5}\left (\frac{\rho}{L}\right)^{5}+\frac{1}{2} \left(\frac{\rho}{L} \right)^{2} \right].
\\
\end{equation}

For the kinetic term function:

\begin{eqnarray}
\label{27}
&&\omega(\rho)=M^{8}\Omega_{4} T_{z} \int \sigma e^{-\varphi_{\infty}/12}\delta(r-\rho)H^{4/3}r^{4}\,dr= \nonumber \\
\\  
&&=M^{4}(ML)^{3}(MT_{z})\Omega_{4} \, 12\left(\frac{2}{3}\right)^{1/3}\left(\frac{\rho}{L}\right)^{4}\left[1+\left(\frac{L}{\rho}\right)^{3}\right]^{4/3}. \nonumber
\end{eqnarray}

And for potential in (\ref{25}) (expression in square brackets includes GH term; gravity constraint was used in receiving it):

\begin{eqnarray}
\label{28}
&&{\widetilde V}(\rho)=M^{8}\Omega_{4} T_{z} \Bigg\{-2\int_{0}^{\rho}\left[\frac{24}{H^{5/8}r^{2}}-\frac{Q_{(4)}^{2}e^{\varphi_{\infty} /2}H^{-1/8}}{H^{5/2}r^{8}}\right]H^{5/8}r^{4}\,dr + 
\\
&&+\int \sigma e^{-\varphi_{\infty} /12} H^{1/48}\delta(r-\rho)H^{5/16}r^{4}\,dr\Bigg\}=4M^{4}(ML)^{3}(MT_{z})\Omega_{4}F\left(\frac{\rho}{L}\right), \nonumber
\end{eqnarray}
where function $F(y)$ is given by the formula:

\begin{eqnarray}
\label{29}
&&F(y)=y^{3}\left[3\left(\frac{2}{3}\right)^{1/3}(1+y^{3})^{1/3}+\frac{3}{2(1+y^{3})}-4\right], \nonumber
\\
&&y=\frac{\rho}{L}, \qquad  y_{0}=\frac{r_{0}}{L}=2^{-1/3},
\end{eqnarray}
value of $y_{0}$ is received from (\ref{18}). It is easy to see that $F(y)$ possesses minimum at $y=y_{0}$ and $F(y_{0})=0$. The same is true for potential ${\tilde V}(\rho)$ (\ref{28}) at $\rho=r_{0}$.

{\it Note 1.} Although Gibbons-Hawking term is a full divergence and we consider compact extra space it would be mistake to discard GH term in (\ref{25}) when radion effective potential is calculated. Direct calculation of the GH term in (\ref{25}), when step-functions reflecting mirror $Z_{2}$-jumps of $b(r)$, $c(r)$, $a(r)$, $\varphi (r)$ are taken into account, shows that it really vanish at the solution of dynamical equations, i.e. at $\rho=r_{0}$. But GH contribution to the radion effective potential is by no means equal to zero when upper limit of integration in (\ref{25}) is changed from $r_{0}$ to arbitrary value $\rho$.

{\it Note 2.} It is impossible to calculate from the action (\ref{1}) the physically meaningful radion effective potential in case dual electric 6-form $F_{6}$ is used in (\ref{1}). Although electric $4$-brane extremal solution is given by the same formulae (\ref{15}), (\ref{18})-(\ref{21}) as a magnetic one, the values of action (\ref{1}), $S_{m}$ and $S_{e}$, calculated at the magnetic and electric fluxbrane solutions as a backgrounds drastically differ. General consistency conditions \cite{Leblond} say that $S_{m}$ must vanish at the solution of dynamical equations but these conditions are not applicable to $S_{e}$ (see Appendix in \cite{Alt06}).

According to (\ref{28}), (\ref{29}) ${\widetilde V}(\rho)\to 0$ at $\rho\to 0$ and ${\widetilde V}(\rho)\to \infty$ at $\rho\to \infty$. However this behavior is of no physical interest since similar behavior possesses the Brans-Dicke field (\ref{26}). To get the physical radion effective potential the low dimension Brans-Dicke effective action (\ref{25}) must be written in the Einstein-frame metric and radion field must be transformed in a way providing the canonical form of its kinetic term. 

Thus let us rescale metric ${\tilde g}_{\mu\nu}$ in the Brans-Dicke action in the RHS of (\ref{25}) to the Einstein-frame metric $g_{\mu\nu}$:

\begin{equation}
\label{30}
{\tilde g}_{\mu\nu}=\frac{M_{\rm Pl}^{2}}{\Phi(\rho)}\,g_{\mu\nu},
\end{equation}
where $M_{\rm Pl}=10^{19}GeV$ is Planck mass.

Effective action (\ref{25}) being expressed as a functional of the Einstein-frame metric $g_{\mu\nu}$ introduced in (\ref{30}) and of the canonical radion field $\psi$ (defined below) takes the standard form:

\begin{equation}
\label{31}
S^{(4)}=\int\left[M_{\rm Pl}^{2}R^{(4)}-(1/2)M_{\rm Pl}^{2}(\nabla \psi)^{2}-\mu^{4}V(\psi)\right]\sqrt{-g^{(4)}}\,d^{(4)}x.
\end{equation}
$\mu$ is a calculable constant of dimensionality of mass - the characteristic of the radion potential, $V(\psi)$ is taken dimensionless for convenience. Also dimensionless (normalized to Planck mass) canonical radion field $\psi(\rho)$ is introduced in (\ref{31}):

\begin{equation}
\label{32}
\psi(\rho)=\frac{1}{L}\int_{r_{0}}^{\rho}\epsilon (\rho)\,d\rho=\int_{y_{0}}^{y}\epsilon (y)\,dy, \qquad y=\frac{\rho}{L},
\end{equation}
here the point (\ref{18}) of stable extremum of the radion effective potential is chosen at $\psi=0$; $y_{0}=2^{-1/3}$; $\epsilon (\rho)$ is expressed through functions $\Phi(\rho)$, $\omega(\rho)$ given in (\ref{26}), (\ref{27}):

\begin{eqnarray}
\label{33}
&&\epsilon^{2}(y)=L^{2}\left[ \frac{\omega(\rho)}{\Phi(\rho)}+3 \left(\frac{1}{\Phi}\frac{d\Phi}{d\rho}\right)^{2}\right]= \nonumber \\
\\
&&=6 \left(\frac{2}{3}\right)^{1/3} y^{4} \left(\frac{y^{5}}{5}+\frac{y^{2}}{2}\right)^{-1}\left(1+\frac{1}{y^{3}}\right)^{4/3}+3y^{8}\left(\frac{y^{5}}{5}+\frac{y^{2}}{2}\right)^{-2}\left(1+\frac{1}{y^{3}}\right)^{2}. \nonumber
\end{eqnarray}

It is seen from (\ref{33}) that in the $\rho \ll L$ ($y \ll 1$) limit, i.e. inside the throat, $\epsilon(y) \sim y^{-1}$ and in the $\rho \gg L$ ($y \gg 1$) limit we have $\epsilon \sim y^{-1/2}$. Hence it follows from (\ref{32}) that in these two limits:

\begin{equation}
\label{34}
\psi (y)=c\cdot \ln y, \qquad c=2 (18^{1/3}+3)^{1/2}, \qquad 0<y=\frac{\rho}{L}\ll 1,
\end{equation}
and

\begin{equation}
\label{35}
\psi (y)=2 (10 \cdot 18^{1/3})^{1/2} \,y^{1/2}, \qquad  1\ll y=\frac{\rho}{L}<\infty.
\end{equation}

Radion potential $\mu^{4}V(\psi)$ in (\ref{31}) is expressed through the auxiliary potential ${\widetilde V}(\rho)$ (\ref{28}) and Brans-Dicke field $\Phi (\rho)$ (\ref{26}):

\begin{eqnarray}
\label{36}
&&\mu^{4}V(\psi)=M_{\rm Pl}^{4} \frac{{\widetilde V}(\rho)}{\Phi(\rho)^{2}}=\frac{M_{\rm Pl}^{4}}{(MT_{z})(ML)^{7}\Omega_{4}}\, K(y(\psi)), \nonumber
\\ \nonumber
\\
&&V(\psi) \equiv K(y(\psi))= \frac{F(y)}{\left(\frac{y^{5}}{5}+\frac{y^{2}}{2}\right)^{2}},
\end{eqnarray}
where function $F(y)$ is given in (\ref{29}) and dependence $y(\psi)$ must be received from (\ref{32}), (\ref{33}). The characteristic density $\mu^{4}$ of the radion effective potential is defined in (\ref{36}). It depends on the period $T_{z}$ of torus $S^{1}$ in the metric (\ref{15}) and on the length of the throat $L$; expression for $ML$ is given in (\ref{19}), value of $MT_{z}$ will be calculated in Sec. 4 below. Because of strong inequality $ML\gg 1$ demanded by the applicability of the low-energy string approximation $\mu^{4}$ in (\ref{36}) proves to be suppressed as compared to the Planck density $M_{\rm Pl}^{4}$. This is important since effective action approach is valid only if radion potential in (\ref{31}) is essentially below the Planck density

\begin{equation}
\label{37}
\mu^{4}V(\psi)\ll M_{\rm Pl}^{4}.
\end{equation}
We'll see in Sec. 4 that although $V(\psi)$ is growing down the throat inequality (\ref{37}) is valid everywhere in the region  of applicability of the low-energy string approximation given by the condition (\ref{17}).

The form of the dimensionless potential $V(\psi)$ (\ref{36}) depends only on the choice of the theory. For the Type IIA supergravity with the co-dimension one local source (choice (\ref{2}) of dimensionalities and coupling constants) $V(\psi)$ is drawn in \cite{Alt06} (Curve "D" at Fig. 1 of \cite{Alt06}). Potential is non-negative, as expected it possesses zero minimum at $\psi=0$ where junction conditions (\ref{11})-(\ref{14}) are valid. To the right of this point $V(\psi)$ increases, reaches maximum and then again falls down to zero at infinity. Thus stable state $\psi=0$ where supposedly our Universe "lives in" is protected from the runaway decompactification by certain potential barrier. This situation is typical for all theories with compactified extra dimensions \cite{Giddings2}. It is not without interest to study in frames of considered model to what extent this "protection" is reliable. But we'll leave this work for future.

Asymptotic behavior of the dimensionless radion potential $V(\psi)$ in the limits $\psi \ll -1$ and $\psi \gg 1$ immediately follow from (\ref{36}) with account of expression (\ref{29}) for $F(y)$ and asymptotic (\ref{34}), (\ref{35}) for $\psi(y)$:

For $\psi \ll -1$:

\begin{equation}
\label{38}
V_{-}(\psi)= (2^{7/3}3^{2/3}-10)e^{-\psi /c} \approx 0,48 \cdot e^{-0,21\cdot\psi},
\end{equation}
where $c$ is given in (\ref{34}).

And at $\psi \gg 1$:

\begin{equation}
\label{39}
V_{+}(\psi)=2^{20}\cdot 3^{5}\cdot 5^{8}\cdot \left(\frac{2}{3}\right)^{1/3}\cdot \psi^{-12} \approx \frac{8,7\cdot 10^{13}}{\psi^{12}}.
\end{equation}

Now we can look shortly at the possibility to apply these results to the description of inflation in the early Universe \cite{Guth}, \cite{Dvali}, \cite{Mukhanov}.
Radion field introduced above hopefully may serve as an inflaton field (cf. \cite[b]{Cline}, \cite{Mazumdar}). We may suppose that initially the "heavy lid" boundary of the extra space was located somewhere deep in the throat ($\rho_{\it in} \ll L$ or $\psi_{\it in} \ll -1$) and after that, obeying the dynamics determined by the action (\ref{31}), it rolls down the exponential asymptotic $V_{-}(\psi)$ (\ref{38}) of the radion potential (\ref{36}) to the steep slope leading to stable brane's position (\ref{18}) ($\psi=0$, see (\ref{32})) at the top of the throat.

The following questions must be answered: Does radion potential $V(\psi)$ (\ref{36}) meet the necessary flatness and slow roll conditions? Can this scenario provide the number of $e$-foldings $N_{e}$ during inflation demanded by the astrophysical observations ($N_{e} \approx 80-100$) \cite{Dvali}-\cite{WMAP}? For the exponentially decreasing potential $V(\psi)\sim e^{-k\psi}$ flatness and slow roll conditions demand $k^{2} \ll 1$ which seemingly is true for $k=0,21$ like in (\ref{38}). The number of $e$-foldings during inflation is given by simple formula \cite{Dvali, Mukhanov} (prime means derivative over $\psi$ which, we remind, is dimensionless - in Planck units):

\begin{equation}
\label{40}
N_{e}=\int_{\psi_{\it in}}^{\psi_{\it fin}}\frac{V(\psi)}{V'(\psi)}\,d\psi = \frac{\psi_{\it fin}- \psi_{\it in}}{k},
\end{equation}
where last equality is received for the exponential potential; $\psi_{\it in}$ and $\psi_{\it fin}$ are the values of the radion (inflaton) field in the beginning and in the end of the inflation. Thus for the value of $k=0,21$ (\ref{38}) it follows from (\ref{40}) that necessary number of $e$-foldings is reached if $\psi_{\it fin}- \psi_{\it in} > 20$.

The end of inflation where reheating begins is expected at the beginning of steep slope of the radion potential. Analyses of the exact analytical expression of $V(\psi)$ (\ref{36}) shows that steep slope begins somewhere at $\psi_{\it fin} \approx -20$. Hence to receive the sufficiently long period of inflation the initial value of the radion (inflaton) field must satisfy $\psi_{\it{in}} \le -40$, i.e. initial position of the "heavy lid" boundary must be sufficiently deep in the throat.

Let us look at the validity of the inequality $\psi_{\it in} \le -40$ from the point of view of applicability of the low-energy string approximation. The permitted values of the isotropic coordinate $r$ must obey inequality $r> r_{\it min}$ where $r_{\it min}$ is determined in (\ref{17}). Corresponding minimal permitted value $\psi_{\it min}$ may be calculated from asymptotic expression (\ref{34}) (where it is taken $y_{\it min}= \rho_{\it min}/L=r_{\it min}/L$). If we express $(ML)$ in (\ref{17}) from (\ref{19}) then the value of $\psi_{\it min}$ is found from (\ref{34}): 

\begin{equation}
\label{41}
\psi_{\it min}=38\cdot \ln k-180-76\cdot \ln g.
\end{equation}
As long as $k$ in (\ref{17}) is of order one and parameter $g\ge 1$ (according to (\ref{19}) it is demanded by the condition of validity of all the approach $ML\gg 1$) this value of $\psi_{\it min}$ is essentially below the value providing the necessary number of e-foldings during inflation ($\psi_{\it in} \approx -40$). Thus permitted length of the throat does not come in conflict with demands of the early inflation.

To be sure that effective action approach of this Section is 
consistent the validity of inequality (\ref{37}) must be established. 
This will be done in subsection {\it 4-c}.

\section{Deformation of the elementary fluxbrane solution}

\vspace{0,5 cm}

{\large\it 4-a. General formulae.}

\vspace{0,5cm}

In this Section equations (\ref{5})-(\ref{8}) will be analized for the case of 
non-zero "Maxwell" field $F_{(2)}$ given in ansatz (\ref{4}) and non-zero small 
constant positive curvature of the Universe, i.e. when $Q_{(2)}\ne 0$, ${\tilde h}\ne 0$ 
in (\ref{5})-(\ref{8}). Let us rewrite the metric of ansatz (\ref{4}) in a form:

\begin{equation}
\label{42}
ds_{(10)}^{2}=b^{2}({\tilde g}_{\mu\nu}dx^{\mu}dx^{\nu}+ Udz^{2})+f^{2}\left(\frac{dr^{2}}{U}+r^{2}d\Omega_{4}^{2}\right),
\end{equation}
this is always possible with a transformation of the isotropic coordinate $r$ in (\ref{4}). 
The non-deformed solution (\ref{15}) of equations (\ref{5})-(\ref{8}) looks for
metric (\ref{42}) as:

\begin{equation}
\label{43}
b=\bar{b}=H^{-3/16}, \quad  U=\bar{U}=1, \quad f=\bar{f}=H^{5/16}, \quad e^{\varphi}=e^{\bar{\varphi}}=e^{\bar{\varphi}_{\infty}}H^{-1/4},
\end{equation}
here we included for convenience the expression (\ref{15}) for the "non-deformed"
dilaton field.

There is well known \cite{Horowitz}-\cite{Aharony} exact Schwarzshild
type bulk solution generalizing 
metric (\ref{15}) in a way of (\ref{42}) where

\begin{equation}
\label{44}
U=U_{\it{Sch}}=1+\frac{const}{r^{3}}
\end{equation}
and $b$, $f$, $\varphi$ are like in (\ref{43}). This solution was used 
in \cite{Alt06} to build the IR end of the throat in the "bolt" point where 
$U_{\it{Sch}}=0$. However, as it was estimated in \cite{Alt06} and was 
shown exactly in \cite{Altsh1} in
6D generalization of the Randall-Sundrum model the value of 
Dark Energy received from the Schwarzshild type deformation of the 
elementary throat-like solution is about 60 orders above the observed value 
$10^{-120}M_{\rm Pl}^{4}$. That is why in \cite{Altsh1} not Schwarzshild type but 
Reissner-Nordsrtrom type deformation of the Randall-Sundrum $AdS$ model was used, 
and it was shown that in this case the calculated value of the Dark Energy 
may be in accordance with observations.

Before starting the analyses of the solution of Eqs. (\ref{5})-(\ref{8}) 
when $Q_{(2)}\ne 0$ and ${\tilde h}\ne 0$ it is worthwile to outline shortly the logic 
of introduction in these models of extremely small positive curvature of 
the Universe. The point is 
that presence of $U(r)\ne const$ in metric (\ref{42}) results in discrepancy 
of the Israel junction conditions at the UV boundary of space-time (discrepancy 
appears since it is 
supposed that energy-momentum tensor of the boundary is isotropic). Because of 
quick decrease of the additional, depending on $r$, term in $U(r)$ with 
increase of $r$ from IR end to UV end of the warped extra space 
this discrepancy at the UV end proves to be quite small. The remedy may be the 
introduction of small non-zero positive curvature of the Universe 
which will give additional term in $U(r)$ reparing Israel junction conditions.
However the decrease with growth of $r$ of the Schwarzshild term (\ref{44}) 
proves to be insufficiently quick and, as it was said above, does not give the 
observed value of the Dark Energy; more satisfactory result may be 
expected when Reissner-Nordstrom type deformation is used.

As for our knowledge the exact solution of Eqs. (\ref{5})-(\ref{8}) 
when $Q_{(2)}\ne 0$, ${\tilde h}\ne 0$ is not found yet. If $Q_{(2)}$, ${\tilde h}$ are small 
as compared to the scales of the non-deformed solution (\ref{15}),
and this is the case under consideration, then induced variations 
of $b(r)$, $f(r)$, $\varphi(r)$ in (\ref{42}) are also small as compared 
to their "non-deformed" 
values (\ref{43}) and may be studied in linear approximation 
of Eqs. (\ref{5})-(\ref{8}). Also small will be the variations of position 
of UV boundary and of the "fine-tuning" condition, i.e. variations of 
$r_{0}$, $Q_{(4)}$ which "non-deformed" values are given in (\ref{18}), 
(\ref{21}). However in the context of the present paper there is no need 
to calculate all these small variations.

The peculiarity of the situation is that change of $U(r)$ in (\ref{42}) does 
not need to be small as 
compared to $U=\bar{U}=1$ of (\ref{43}). To receive $U(r)$ it is 
sufficient to subtract equations (\ref{5}), (\ref{6}) where in accordance 
with (\ref{42}) we put $c^{2}=U b^{2}$, $N^{2}=f^{2}/U$, $a=rf$. The resulting 
equation for $U(r)$ looks as follows:

\begin{equation}
\label{45}
U''+U'\left(5\frac{b'}{b}+3\frac{f'}{f}+\frac{4}{r}\right)=-2f^{2}\,\frac{e^{-3\varphi /2}Q_{(2)}^{2}}{2b^{8}f^{8}r^{8}}-f^{2}\,\frac{6{\tilde h}^{2}}{b^{2}}.
\end{equation}

Since $Q_{(2)}$, ${\tilde h}$ in the RHS of (\ref{45}) are supposed to be small, the other functions 
in (\ref{45}) ($b(r)$, $f(r)$, $\varphi(r)$) may be taken in zero 
approximation. Substitution of their expressions (\ref{43}) 
into (\ref{45}) gives:

\begin{equation}
\label{46}
U''+\frac{4}{r}U'=-\frac{Q_{(2)}^{2}e^{-3\varphi_{\infty}/2}}{r^{8}}-6{\tilde h}^{2}\left(1+\frac{L^{3}}{r^{3}}\right).
\end{equation}
The free solution of (\ref{46}) is, as expected, the Schwarzshild potential 
(\ref{44}); in what follows we shall discard this term of $U(r)$ for the reasons 
explained above in this Section.

Substraction of junction conditions (\ref{11}), (\ref{12}) with 
account $c^{2}=Ub^{2}$ gives simple condition of their consistency:

\begin{equation}
\label{47}
U'(r_{0})=0.
\end{equation}
Strictly speaking (\ref{47}) must be valid at the location of 
the $Z_{2}$-symmetric UV boundary slightly shifted from its "non-deformed"
position (\ref{18}). But in the lowest approximation we may take 
in (\ref{47}) $r_{0}=L/2^{1/3}$ given in (\ref{18}). From (\ref{47}) the value 
of ${\tilde h}$ will be determined.

\vspace{0,5 cm}

{\large\it 4-b. Case $F_{(2)}\ne 0$, ${\tilde h}=0$. Determination of modulus $T_{z}$.}

\vspace{0,5cm}

As it will be seen the value of ${\tilde h}$ determined from condition (\ref{47}) is 
extremely small, hence at the IR end of the throat and practically everywhere 
inside the throat the ${\tilde h}$-term is essentially below the $Q_{(2)}$-term 
in the RHS of (\ref{46}). Thus let us first put down the solution of 
equation (\ref{46}) in case ${\tilde h}=0$:
 
\begin{equation}
\label{48}
U=1-\left(\frac{l}{r}\right)^{6}, \qquad l^{6} \equiv \frac{Q_{(2)}^{2}e^{-3\varphi_{\infty}/2}}{18},
\end{equation}
it is supposed that $l\ll L$, which means that deformation (\ref{48}) of 
the elementary solution (\ref{15}) is small; we remind that Schwarzshild 
term $\sim r^{-3}$ (see (\ref{44})) is deliberately omitted in (\ref{48}). 
Metric (\ref{42}) 
with $U(r)$ given in (\ref{48}) is a Euclidian "time" 
version of the Reissner-Nordstrom generalization of the elementary throat-like 
solution. 

The "bolt" point $r=l$ where $U=0$ is the IR end of the throat; it is 
topologically equivalent to the pole of 2-sphere 
\cite{Hawking}-\cite{Aghababaie}. Space-time (\ref{42}) may possess 
conical singularity at this point in case "matter trapping" co-dimension 
two IR brane is placed there (see e.g. \cite{Aghababaie, Carroll, Rubakov2}). 
This will produce deficit angle $\delta_{d}$ depending on tension of 
the IR brane which will influence the value of period $T_{z}$ of the Euclidian 
"time" $S^{1}$ calculated from (\ref{42}), (\ref{48}). We shall not consider this 
option in the present paper and postulate that $\delta_{d}=1$, 
hence IR end of the throat is supposed to be  
smooth. Then taking the zero-order dependences (\ref{43}) for $b(r)$, $f(r)$ 
in metric (\ref{42}) and with account of (\ref{48}) for $U(r)$ the following 
expression for period of torus $S^{1}$ of space-time (\ref{15}) (or (\ref{42})) 
is received:

\begin{equation}
\label{49}
T_{z}=\frac{2\pi}{3}\,H^{1/2} \, l \approx \frac{2\pi}{3} \, L \, \left(\frac{L}{l}\right)^{1/2},
\end{equation}
where $H(r)$ is given in (\ref{15}) and it is taken at $r=l$, last approximate 
equality is valid since $l \ll L$. Thus 
period $T_{z}$ of the extra torus of space-time (\ref{42}) is not an 
arbitrary modulus of the solution but is determined by (\ref{49}) through 
the characteristic lengths of the throat $L$, $l$. From (\ref{49}) 
it follows that $T_{z} \gg L$.

Substitution of modulus (\ref{49}) in expression (\ref{26}) 
for Brans-Dicke field $\Phi$, with account of (\ref{18}), (\ref{19}),
gives the important quantity, entering formulae for hierarchies 
in Sec. 5, as a function of $l$ and parameter $g$ (\ref{20}):

\begin{equation}
\label{50}
\frac{M}{\sqrt{\Phi(r_{0})}}=10^{-4}\, g^{-3} \, \left(\frac{l}{L}\right)^{1/4},
\end{equation}
where coefficient $10^{-4}$ absorbs numbers of formulae (18) for $r_{0}$, 
(\ref{19}) for $ML$, (\ref{26}) for $\Phi$ and (\ref{49}) for $T_{z}$, 
including the value of volume of 4-sphere of unit 
radius $\Omega_{4}=8\pi^{2}/3$ in (\ref{26}).

\vspace{0,5cm}

{\large\it 4-c. Consistency of the effective action approach.}

\vspace{0,5cm}

Now when we determined the modulus $T_{z}$ (\ref{49}) it is possible to 
check up the enaquality (\ref{37}) which is the condition of 
applicability of the low-dimension effective action approach of Sec. 3. 
Since potential (\ref{36}) grows down the throat it is sufficient to 
verify (\ref{37}) at the IR end of the throat, i.e. at $\rho = l$. 

Substitution in (\ref{37}) of asymptotic expressions (\ref{38}), (\ref{34})
for $V(\psi)$, $\psi(\rho)$ (and with account of formulae (\ref{19}), (\ref{49})
for $ML$, $MT_{z}$ entering in the definition of $\mu^{4}$ in (\ref{36})) 
inequality (\ref{37}) at $\rho=l$ is expressed through location $l$ of the IR 
end of the throat and parameter $g$ defined in (\ref{20}):

\begin{equation}
\label{51}
\frac{\mu^{4}}{M_{\rm Pl}^{4}}\,V(\psi(l))=6\cdot 10^{-11}\,g^{-8}\,\left(\frac{L}{l}\right)^{1/2} < 1.
\end{equation}

The location of the IR end of the throat must meet the 
demand $l>r_{\it min}$ of validity of the low-energy action (\ref{1}), 
where $r_{\it min}$ is estimated in (\ref{17}). It is interesting to note that 
even at this depth inequality (\ref{51}) remains to be valid. More of that: its 
validity does not depend on the value of parameter $g$ which drops out from 
the expression (\ref{51}). In fact substitution of $l=r_{\it min}$ 
in (\ref{51}) with account of expression (\ref{17}) for $r_{\it min}$ gives:

\begin{equation}
\label{52}
\frac{\mu^{4}}{M_{\rm Pl}^{4}}\,V(\psi(r_{\it min}))= \frac{10^{-2}}{k^{4}} < 1,
\end{equation}
this inequality is valid since coefficient $k$ introduced 
in (\ref{17}) is supposed to be of order of one.

\vspace{0,5 cm}

{\large\it 4-d. Adjustment of Israel conditions and determination of ${\tilde h}$.}

\vspace{0,5cm}

Condition (\ref{47}) of consistency of the Israel junction equations for 
subspaces $M_{(3+1)}$ and $S^{1}$ of the boundary of space-time (\ref{42})
can not be fulfilled for $U(r)$ given in (\ref{48}). To repair Israel 
conditions the necessary anisotropy of the energy-momentum tensor of the 
boundary was introduced in \cite{Louko}, \cite{Aghababaie} in 6D model. 
But perhaps
it would be more natural to escape the arbitrary modifications of the action 
of local source in (\ref{1}) and to resolve the problem with introduction of 
small positive curvature of space-time $M_{(3+1)}$ \cite{Altsh1}. We shall go 
this way. Thus taking ${\tilde h}\ne 0$ in the RHS of (\ref{46}) and with 
account of (\ref{48}) the following expression for $U(r)$ 
is received from equation (\ref{46}):

\begin{equation}
\label{53}
U=1-\left(\frac{l}{r}\right)^{6}-\frac{3}{5}{\tilde h}^{2}r^{2}+\frac{3{\tilde h}^{2}L^{3}}{r}.
\end{equation}
Then ${\tilde h}$ is immediately determined from (\ref{47}) 
(where $r_{0}=L/2^{1/3}$ (\ref{18})):

\begin{equation}
\label{54}
{\tilde h}=\sqrt{\frac{2^{5/3}\cdot 5}{3}}\,\frac{1}{L}\, \left(\frac{l}{L}\right)^{3}.
\end{equation}

Condition $U=0$ gives location of the IR end of the throat. Presence 
of the ${\tilde h}$-terms in expression for $U(r)$ (\ref{53}) will make a 
shift of this position from the value $r=l$ detremined from (\ref{48}). This 
shift is however extremely small since it follows from (\ref{54}) that 
at $r=l$ the main (second one) ${\tilde h}$-term in the RHS of (\ref{53}) 
is suppressed by the factor $(l/L)^{5}$ as compared to the $Q_{(2)}$-term. 
Since $l\ll L$ we may consider $r=l$ the location of the IR end of the throat.
Actually "curvature" ${\tilde h}$-terms may be neglected in (\ref{53}), as well 
as in the RHS of (\ref{46}), practically everywhere inside the throat; they 
become comparable with "Maxwell" $Q_{(2)}$-terms only in vicinity of 
the top of the throat $r\cong L$, although both remain quite small there.

Auxiliary "Hubble constant" ${\tilde h}$ (\ref{54}) 
characterises the curvature ${\widetilde R}^{(4)}$ 
of the manifold $M_{(3+1)}$ (see (\ref{9})). To receive the 
observed rate of acceleration of the Universe $h$
we must rescale ${\widetilde R}^{(4)}$ with transformation (\ref{30}) 
to the Einstein-frame curvature $R^{(4)}$. This will be done 
in the next Section.

\section{Calculation of the mass scale hierarchy and of the "acceleration hierarchy"}

{\large\it 5-a. Formula for mass scale hierarchy.}

\vspace{0,5cm}

Following the Randall and Sundrum approach \cite{Randall} we take 
mass parameters of matter action written in the primordial metric 
of the action (\ref{1}) equal to the fundamental scale $M$. And it is 
conventionally supposed that massive matter of Standard Model is concentrated 
near the IR end of the strongly warped space-time - let it be because of
trapping at the IR brane or because of pure gravitational accretion to the IR end. 
Then, in case warped throat-like solution (\ref{15}) is considered, 
mass of the visible matter is decreased as compared to $M$ by the value 
of warp factor $H^{-3/16}$ at $r=r_{\it IR}$:

\begin{equation}
\label{55}
M \to M\,H^{-3/16}(r_{\it IR}) \approx M\, \left(\frac{r_{\it IR}}{L}\right)^{9/16},
\end{equation}
where it was taken into account that $H \approx (L/r)^{3}$ at $r\ll L$.

In previous Section the IR end of the throat was built as a "bolt" 
point $r=l$ of the deformed metric (\ref{42}), (\ref{48}). In what 
follows we shall put

\begin{equation}
\label{56}
r_{\it IR}=l > r_{\it min},
\end{equation}
where $r_{\it min}$ (\ref{17}) is the point in the depth of the throat where 
effective low-energy string induced action (\ref{1}) is not valid any more.

To receive the observed electro-weak 
scale $m$ it is necessary to write down the effective matter action in 
lower dimensions in the Einstein-frame metric $g_{\mu\nu}$ 
introduced in (\ref{30}). Brans-Dicke field 
$\Phi(\rho)$ in (\ref{30}) must be taken at $\rho=r_{0}$ (\ref{18}) i.e. 
in the minimum of the radion effective potential (\ref{28}) (or 
equivalently in the minimum 
of potential (\ref{36}) at $\psi=0$) where supposedly our Universe 
is stabilized after inflation and reheating. 

In calculating mass scale hierarchy it is possible to use 
expression (\ref{26}) for $\Phi (r_{0})$ recived by integration out of extra 
coordinates in action (\ref{1}), (\ref{2}) when non-deformed 
solution (\ref{15}) is taken as a background. The only 
impact of deformation upon these calculations is dynamical fixation of the 
IR end of the throat at $r=l$ and determination of period of 
torus $T_{z}$ (\ref{49}) performed in Sec. 4. Thus from (\ref{55}), (\ref{30}), (\ref{50}) 
the following expression is received for the mass scale hierarchy as a 
function of location of the IR end of the throat $l$ and parameter $g$ (\ref{20}) 
of the fluxbrane solution:

\begin{equation}
\label{57}
\frac{m}{M_{\rm Pl}}=H^{-3/16}\,\frac{M}{\sqrt{\Phi(r_{0})}}= 10^{-4} \cdot g^{-3} \cdot \left(\frac{l}{L}\right)^{13/16}.
\end{equation}

In case $g=1$ to get the observed value of mass hierarchy it is necessary to place IR end of the throat sufficiently 
deeply: $l/L\approx 10^{-16}$. This value of $l$ practically coicides with the 
limit of validity of the low-energy approximation $r_{\it min}$ (\ref{17}).
For $g>1$ situation becomes less dangerous.

For the limiting depth of the throat, i.e. when we substitute in (\ref{57}) 
$l=r_{\it min}$ from (\ref{17}) (where formula (\ref{19}) for $ML$ is used) 
expression (\ref{57}) reads:

\begin{equation}
\label{58}
\frac{m}{M_{\rm Pl}}= 10^{-17}\, k^{13/2} \,  g^{-16}.
\end{equation}
It is seen that (\ref{58}) gives the observed value of mass scale hierarchy 
$m/M_{\rm Pl}=10^{-16}$ for 
$g\approx 1$, $k\approx 1$. Of course this game in numbers should not be taken 
seriously since the RHS of (\ref{58}) is strongly dependent on free parameters $k$, $g$. It is 
interesting however that wishful big value of mass hierarchy may be received without 
introduction of big numbers "by hand".

Big number $10^{17}$ in (\ref{58}) appeared here "from nothing", i.e. from 
coefficient 10,5 in (\ref{19}) (in a general case equal 
to $4(n-1)[\Delta/2(n-1)]^{1/(n-1)}$, see (\ref{26}) of \cite{Alt06}, 
$\Delta$ is given in (\ref{9}) of \cite{Alt06}) and from the exponent 16 in 
(\ref{17}) (in a general case equal to $\Delta/ \alpha^{2}$). Here we 
have $n=4$, $\Delta=4$, $\alpha = 1/2$.

Physically expression (\ref{58}) for the mass scale hierarchy follows from the 
"bold" hypothesis of \cite{Alt06}, \cite[c]{Altsh} that SM resides at the 
brink of existence of the target space-time, i.e. that massive 
matter falls down the very "bottom" of the throat and concentrates there 
being stopped by unknown higher-curvature terms not included in the low-energy 
action (\ref{1}). In any case the unambigous result of the paper not depending 
on these speculations is given by expression (\ref{57}) for the mass scale hierarchy.

\vspace{0,5cm}

{\large\it 5-b. Values of the acceleration rate and of Dark Energy.}

\vspace{0,5cm}

In Subsection {\it 4-d} expression (\ref{54}) for the auxiliary "Hubble 
constant" $\tilde h$ was deduced from the Israel junction conditions at 
the UV boundary of the throat. To find the observed rate of acceleration 
of the Universe $h$ (equal to $10^{-60}M_{\rm Pl}$ according to the observations) 
it is necessary to perform scale transformation (\ref{30}) taken at the 
point $\rho=r_{0}$ of extremum of the radion potential 
(like it was done for $m$ in expression (\ref{57}) above):

\begin{equation}
\label{59}
\frac{h}{M_{\rm Pl}}=\frac{\tilde h}{M} \, \frac{M}{\sqrt{\Phi(r_{0})}} = 10^{-5} \cdot g^{-4} \cdot \left(\frac{l}{L}\right)^{13/4}.
\end{equation}
Last equality was received from (\ref{54}), (\ref{50}), (\ref{19}).
 
Finally it is instructive to express "acceleration hierarchy" $h/M_{\rm Pl}$ 
not through $l/L$ but through the value of mass hierarchy $m/M_{\rm Pl}$ (\ref{57}):

\begin{equation}
\label{60}
\frac{h}{M_{\rm Pl}}=10^{11} \, g^{8} \left(\frac{m}{M_{\rm Pl}}\right)^{4}.
\end{equation}
The simple, 4-th power, dependence of two hierarchies is a real gift after many 
of cumbersome exponents above.
                              
Dark Energy $\rho_{D.E.}$ responsible for acceleration 
of the Universe $h$ (\ref{60}) is equal to:

\begin{equation}
\label{61}
\rho_{D.E.}=\frac{1}{2} \, M_{\rm Pl}^{2}\, R^{(3+1)}=6h^{2}\, M_{\rm Pl}^{2}=6\cdot 10^{22}\, g^{16}\, \frac{m^{8}}{M_{\rm Pl}^{4}},
\end{equation}
It is seen that in case parameter $g=1$ ($g$ 
is defined in (\ref{20})) the observed value of the Dark Energy 
$10^{-120}M_{\rm Pl}^{4}$ is received from (\ref{61}) for $m=10 \, GeV$.

In the limiting depth of the throat substitution 
of $r_{\it IR}=l=r_{\it min}$ (\ref{17}) into (\ref{59}) gives:

\begin{equation}
\label{62}
\frac{h}{M_{\rm Pl}}=10^{-57} \, g^{-56} \, k^{26}.
\end{equation}
The drawback of this expression, like of (\ref{58}) for the mass scale hierarchy, 
is strong dependence of the RHS on the values of arbitrary parameters 
of order one. We outline however that 
possibly main result of the paper $\rho_{D.E.} \sim G_{N}^{2}m^{8}$ (\ref{61})
($G_{N}=M_{\rm Pl}^{-2}$ is Newton's constant) does not depend on the "bold" 
hypothesis described in the end of previous subsection.

\vspace{0,5cm}

{\large\it 5-c. $\rho_{D.E.}$ as a value of radion potential in its extremum.}

\vspace{0,5cm}

In Sec. 3 the exact analytical form of the radion effective potential was 
received for the non-deformed background solution; it was shown that potential 
possesses zero minimum at the 
value of radion field where all dynamical equations, including junction conditions 
(\ref{11})-(\ref{14}), are fulfilled. To repeat the same calculations for the 
deformed background (\ref{42}) is not a simple task, the more so that we do not 
know the exact bulk solution when $F_{(2)}\ne 0$ in anzats (\ref{4}) and 
when curvature of the manifold $M_{(3+1)}$ is not equal to zero. It is possible 
to show however that deformation of the background will result in tiny shift
of the value of potential $\mu^{4}\,V_{\it extr}$ (\ref{36}) in its extremum 
from zero to the value equal the 
Dark Energy $\rho_{D.E.}$ (\ref{61}).

The tool for calculation of $\mu^{4}\,V_{\it extr}$ is the general  
consistency condition of paper \cite{Leblond} which is valid at the solution 
of the dynamical equations. In our case expression (\ref{12}) of \cite{Leblond}
written when parameter $\alpha_{[49]}$ of this paper is taken equal to 
$p=3$ gives:
                                     
\begin{equation}
\label{63}
\oint b^{4}~(T^{m}_{m}-T^{\mu}_{\mu})=-2 \oint b^{2}{\widetilde R}^{(3+1)},
\end{equation}
here $T^{m}_{m}$, $T^{\mu}_{\mu}$ are traces of the energy-momentum tensor 
of matter fields ($F_{(4)}$, $F_{(2)}$, $\varphi$) in (\ref{1}) in internal 
subspaces and in 4 dimensions correspondingly; $b$ is warp factor in metric (\ref{4}) 
($W$ in \cite{Leblond}); $\oint$ symbolizes the integration over compact 
internal space which is deciphered in the LHS of (\ref{25}) where upper limit 
of integration $\rho$ is to be taken the value $\rho=r_{0}$ (\ref{18}) 
determined by the dynamical 
jump conditions (\ref{11})-(\ref{14}). ${\widetilde R}^{(3+1)}$ is curvature 
of the manifold $M_{(3+1)}$ of space-time (\ref{4}) which is equal to 
zero for the non-deformed space-time (\ref{15}) and 
is equal to $12{\tilde h}^{2}$ in case deformed metric (\ref{42}) is considered.

Also, as it was shown in Appendix in \cite{Alt06}, in case 
form-fields of the action (\ref{1}) "live" only in internal space 
the combination of components of the energy-momentum tensor in the LHS of 
(\ref{63}) is proportional to Lagrangian $\rm L$ of the action (\ref{1}) calculated 
at the solution of dynamical equations:

\begin{equation}
\label{64}
T^{m}_{m}-T^{\mu}_{\mu}=-4{\rm L}. 
\end{equation}

Hence from (\ref{63}), (\ref{64}) it follows that at $\rho=r_{0}$ in (\ref{25}):

\begin{equation}
\label{65}
\oint b^{4} {\rm L}=\frac{1}{2}~\Phi_{\it extr}~{\widetilde R}^{(3+1)}.
\end{equation}
We took into account here that $\oint b^{2}=\Phi_{\it extr}$ - value of Brans-Dicke 
field in (\ref{25}) at the point of extremum of potential.  

According to definition (\ref{25}) effective action in 4 dimensions 
in the point of extremum where radion field is constant is equal to:

\begin{equation}
\label{66}
\oint b^{4} {\rm L}=\Phi_{\it extr}{\widetilde R}^{(3+1)}-{\widetilde V}_{\it extr}={\widetilde V}_{\it extr},
\end{equation}
last equality is valid because value 
of action is calculated at the solution of Einstein equations in 4 dimensions.

Thus finally from (\ref{65}), (\ref{66}) it 
follows: $\oint b^{4} {\rm L}={\widetilde V}_{\it extr}$. 
The same is true for the extremal value of potential in the Einstein frame 
action (\ref{31}):

\begin{equation}
\label{67}
\mu^{4}\,V_{\it extr}=\frac{1}{2}\,M_{\rm Pl}^{2}\,R^{(3+1)}=6h^{2}\,M_{\rm Pl}^{2}=\rho_{D.E.},
\end{equation}
$\rho_{D.E.}$ see in (\ref{61}).

This rather strong result does not depend on details of the solution and 
follows from the consistency condition (\ref{63}) in case proportionality 
(\ref{64}) is fulfilled. To check up (\ref{64}) is a simple task. 
Whereas (\ref{63}) was received in \cite{Leblond} after certain integral 
of full divergence over compact internal space was put equal to zero. 
And in this point the special caution is demanded as it was noted 
in \cite{Leblond} as well. It would be important to check up with 
direct calculation consistency condition (\ref{63}) for the space-time 
of type (\ref{42}) 
with the "bolt" point (where $U=0$) topologically equivalent to 2-sphere.

\section{Conclusion}

\qquad  Paper presents three apparently physically interesting results:

1) Exact expression (\ref{36}) for the scalar field 
potential $\mu^{4}\,V(\psi)$ in the effective action (\ref{31}) calculated 
for the non-deformed fluxbrane solution as a background. Asymptotic 
(\ref{38}) of potential describes slow-roll inflation, potential 
possesses steep 
slope for reheating and zero minimum where matter dominating evolution of 
the Universe begins.

2) Formula (\ref{67}) for the tiny positive deviation (seen today 
as Dark Energy $\rho_{D.E.}$ (\ref{61})) of the extremal value 
of the radion effective potential calculated for the "deformed" background.

3) Expressions (\ref{57}) for mass scale hierarchy $m/M_{\rm Pl}$, (\ref{59}) for 
"acceleration hierarchy" $h/M_{\rm Pl}$ and their relation (\ref{60}) which 
gives non-trivial dependence $\rho_{D.E.} \sim G_{N}^{2}m^{8}$ (\ref{61}), 
$G_{N}$ is Newton's constant. This dependence is a progress as compared to Zeldovich 
"numerology" where $\rho_{D.E.} \sim G_{N}m^{6}$ \cite{Zeldovich}.

Also it was demonstrated that under natural additional hypothesis (that SM 
resides at the boarder of space-time where low-energy string approximation 
stops to be valid) big numbers ($10^{17}$, $10^{57}$ in (\ref{58}), 
(\ref{62})) may be received "from nothing", i.e. from dimensionalities 
$D=10$, $n=4$ and the value of the 4-form-dilaton coupling 
constant $\alpha=1/2$.

All quantative results of the paper depend solely on the choice of 
the theory - the Type IIA supergravity in this 
paper. It would be interesting to trace the logic of the paper for some other 
theories.

Since canonical radion field $\psi$ is associated with position $\rho (x)$ 
of the UV boundary terminating the throat (see (\ref{22}), (\ref{32})) it 
would be interesting to find the description of the non-trivial effective 
dynamics in 4 dimensions given by action (\ref{31}) with potential 
(\ref{36}) in the language of heavy boundary moving in higher dimensional 
non-stationary background formed with account of gravitational back-reaction 
of this $Z_{2}$-symmetric co-dimension one local source.

The basic difficulty of the approach of the paper is the lack of physical 
grounds for the very appearance of the UV boundary surface of the throat 
and for the choice of its dynamics. The simplest Nambu-Goto choice taken 
in the action (\ref{1}) is crucial for the calculations of the paper. But 
the "simplest" does not mean "well grounded".

\section*{Acknowledgements} Author is grateful for plural discussions to M.Z. Iofa and to participants of the Quantum Field Theory Seminar of the Theoretical Physics Department, Lebedev Physical Institute. This work was partially supported by the grant LSS-4401.2006.2

\end{document}